\documentclass[structabstract]{aa}  

\usepackage{graphicx}
\usepackage{natbib}
\usepackage{bm}
\begin{document}
   \title{On the Stability of Elliptical Vortices in Accretion Discs}
   \titlerunning{Stability of Elliptical Vortices}


   \author{Geoffroy Lesur and John C. B. Papaloizou}
   \authorrunning{G. Lesur and J. C. B. Papaloizou}
   \institute{Department of Applied Mathematics and Theoretical Physics, University of Cambridge, Centre for Mathematical Sciences,
Wilberforce Road, Cambridge CB3 0WA, UK \\
              \email{g.lesur@damtp.cam.ac.uk}
                     }

   \date{Received date / Accepted date}

 
  \abstract
   {The existence of large-scale and long-lived 2D vortices in accretion discs has been debated for more than a decade. They appear spontaneously in several 2D disc simulations and they are known to accelerate planetesimal formation through a dust trapping process. In some cases, these vortices 
   may even lead to an efficient way to transport angular momentum in protoplanetary discs when MHD instabilities are inoperative. However, the
  issue of the stability of these structures to the imposition of 3D disturbances is still not fully understood,  and it casts doubts on their
     long term survival}
   {We present new results on the 3D stability of elliptical vortices embedded in accretion discs, based on a linear analysis and several non-linear simulations.}
   {We introduce a simple steady 2D vortex model which is a non-linear solution of the equations of motion, and we show that its core is made of elliptical streamlines. We then derive the linearised equations governing the 3D perturbations in the core of this vortex, and we show that they
    can be reduced to a Floquet problem. We solve this problem numerically in the astrophysical regime, including a simplified model to take into account vertical stratification effects. We present several analytical limits for which the mechanism responsible for  instability can be explained. Finally, we compare the results of the linear analysis to some high resolution numerical simulations obtained with spectral and finite difference methods. A discussion is provided, emphasising the astrophysical consequences of our findings for the dynamics of vortices.}
   {We show that most anticyclonic vortices are unstable due to a resonance between the turnover time and the local epicyclic oscillation period. A small linearly stable
  domain is found for vortex cores with an aspect-ratio of around 5.
  However, our simulations show that it is only the vortex core that is stable, with the instability still appearing  on the vortex boundary. In addition, we find numerically that results obtained under the assumption of incompressibility are not affected by 
the introduction of a moderate compressibility. Finally, we show that a strong vertical stratification does not create any additional stable domain of aspect ratio, but it significantly reduces growth rates for relatively weak (and therefore elongated) vortices.}
   {Elliptical vortices are always unstable, whatever the horizontal or vertical aspect-ratio is. The instability can however be weak and is often found at small scales, making it difficult to detect in low-order finite-difference simulations.}


   \maketitle
%

\section{Introduction}
The existence of 2D long-lived vortices in accretion discs was first proposed by \cite{W44} in an outmoded model of planet formation. This idea was revived by \cite{BS95} to accelerate planetesimals formation by a dust trapping process. This kind of vortex is often observed in 2D simulations of discs \citep[see e.g.][]{GL99,UR04,JG05b,BTC07}, since 2D turbulence is known to generate an inverse cascade of energy leading to large 2D vortices \citep{O49}. Vortices may also be generated by 2D instabilities such as Rossby wave instabilities \citep{LLC99} or baroclinic instabilities \citep{KB03,PJS07}, although the latter is still a matter of ongoing  debate \citep{JG05,JG06}. These vortices may play at least two important roles regarding accretion disc dynamics. First, they could lead to an efficient angular momentum transport process in regions in which the magneto-rotational instability \citep{BH98} doesn't operate, such as in dead zones \citep{G96}. Second, they are a very efficient way to accelerate the planetesimal formation process in protoplanetary discs \citep{JAB04}. However, the stability of these vortices when small 3D disturbances are imposed is largely unknown.

In the astrophysics community, this issue has been investigated mainly numerically. \cite{SSG06} examined the formation of 2D vortices starting from 2D turbulence in fully compressible simulations. According to their results, a small 3D noise added to their initialy 2D configuration destroys the coherent vortices in a few orbits, relaxing the flow to its laminar state. \cite{BM05} also computed the evolution of 3D vortices using an anelastic code incorporating vertical stratification.
As \cite{SSG06}, they found that midplane vortices were destroyed by 3D perturbations. However, they also showed that off-midplane vortices could  survive for several hundreds of orbits, leading to the possibility of a stabilizing effect due to the stratification

It is often assumed that these vortices are unstable because 
of the elliptical instability. 
The elliptical instability is a parametric
instability appearing when a multiple of
the vortex turnover frequency matches an
inertial wave frequency, leading to a positive resonance. 
It is observed when the backgound flow follows closed streamlines, 
and being localized on individual streamlines is a \emph{local} instability 
(in particular it doesn't need to involve the vortex boundaries). 
This instability was first found numerically by \cite{P86} and described 
using \cite{CC86} solutions by \cite{B86} for pure elliptical flows.
The rotating case was studied by \cite{C89}, who showed that anticyclonic elliptical flows 
can be stable for some rotation rates. Interested readers may consult \cite{K02} for a 
more extensive discussion of the elliptical instability and its development in fluid mechanics.

In the present paper, we investigate the elliptical instability in the context of accretion disc vortices. 
We first present a steady 2D vortex model, which is a non-linear solution of the local disc equations. 
We then present the linearised equations governing  3D perturbations inside the vortex. 
A criterion for the instability is derived from these equations and a physical understanding of 
the mechanism responsible for the instability is provided. 
We briefly extend these results to a simplified stratified case, 
and we compare our findings to fully non-linear simulations of accretion discs vortices.
 Finally, we provide a discussion and a comparison with previous work.

\section{\label{model_section}Local model of an elliptical vortex.}
\subsection{Shearing-sheet model}
In the following, we will assume a local model for the accretion disc, following the shearing-sheet approximation. The reader may consult \cite{HGB95}, \cite{B03} and \cite{RU08} for an extensive discussion of the properties and limitations of this model. As a simplification, we will assume the flow is incompressible, consistently with the small shearing box model \citep{RU08}. The shearing box equations are found by considering a Cartesian box centred at $r=R_0$, rotating with the disc at angular velocity $\Omega=\Omega(R_0)$.
We define $R_0\phi \rightarrow x$ and $r-R_0 \rightarrow -y$ for consistency with the standard notation for plane Couette flows \citep[e.g. ][]{DR81}. Note that this definition differs from the standard notation used in shearing boxes \citep{HGB95} with $x\rightarrow -y_\mathrm{SB}$, $y\rightarrow x_\mathrm{SB}$ and $z\rightarrow z_\mathrm{SB}$. In this rotating frame, one  obtains the following set of governing equations
\begin{eqnarray}
\label{motiongeneral} \partial_t \bm{u}+\bm{\nabla\cdot} (\bm{u\otimes u})&=&-\bm{\nabla} \Pi
-2\bm{\Omega \times u}+2\Omega S y \bm{e_y},\\
\label{divv} \bm{\nabla \cdot u}&=&0.
\end{eqnarray}
In these equations, we have defined the mean shear $S=-r\partial_r \Omega$, which is set to $S=(3/2)\Omega$ for a Keplerian disc. The generalised pressure $\Pi= P/\rho_0$ is calculated solving a Poisson equation with the incompressibility condition. One can check easely that the velocity field $\bm{u}=Sy\bm{e_x}$ is a steady solution of these equations.

\subsection{The \cite{K81} solution\label{kidavortex}}
We want to study the stability of a steady elliptical vortex embedded in the sheared flow described in the previous subsection. Since we are looking for 2D solutions in the $(x,y)$ plane, we can omit the Coriolis force as it will only change the pressure distribution. We define this vortex by an elliptical patch of constant vorticity $\omega_t = -S + \omega_v$ (the ``core'') where $-S$ is the background flow vorticity and $\omega_v$ is the vorticity of the vortex itself. Outside of this core, the vorticity is assumed to be $\omega_t = -S$, extending to infinity. According to \cite{K81} (Eq. 2.9), such a vortex is steady if the semi-major axis is aligned with $x$ and if its vorticity satisfies
\begin{equation}
\frac{\omega_v}{S}=-\frac{1}{\chi}\Big(\frac{\chi+1}{\chi-1}\Big),
\end{equation}
where we have defined the vortex aspect-ratio $\chi = a/b$, $a$ and $b$ being respectively the vortex semi-major and semi-minor axis. One deduces from this result that only anticyclonic vortices ($\Omega$ and $\omega_v$ having opposite sign) are stable in Keplerian flows. Since this solution is steady, no streamline goes through the core boundaries, and the streamlines \emph{inside} the core have to be elliptical, with the same aspect-ratio as the vortex core.

Thanks to this property, we can write the velocity field in the vortex core, assuming it's centered on $x=y=0$, as
\begin{eqnarray}
\label{v2Dx} u_x^0 &= & S\frac{1}{\chi-1} \chi y,\\
\label{v2Dy} u_y^0 &= & -S \frac{1}{(\chi-1)} \frac{1}{\chi}x.
\end{eqnarray}
This solution can be written in the simpler form
\begin{equation}
u_i^0 = S A_{ij} x_j,
\end{equation}
defining
\begin{equation}
\bm{A}= \frac{1}{\chi-1}\left ( \begin{array}{ccc}
0     &  \chi   &   0 \\
-\chi^{-1} & 0     &   0 \\
0      &   0    &    0 \\
\end{array} \right).
\end{equation}

\subsection{Explicit solution}
A complete solution for the velocity field can be found defining the streamfunction $\psi(x,y)$ so that $u_x=-\partial_y \psi$ and $u_y=\partial_x \psi$. This streamfunction satisfies:

\begin{equation}
\label{poisson}\Delta \psi=\left\{ \begin{array}{ll}
-S+\omega_\nu& \textrm{inside the core,}\\
-S&\textrm{outside.}\\
\end{array} \right.
\end{equation}
One solves these equations in elliptical coordinates, defining $(\mu,\nu)$ by
\begin{eqnarray}
x&=&f\cosh (\mu) \cos (\nu),\\
y&=&f\sinh (\mu) \sin (\nu),
\end{eqnarray}
with $f=a\sqrt{(\chi^2-1)/\chi^2}$. In these coordinates, the core boundary is found at $\mu=\mu_0$ with $\tanh(\mu_0)=\chi^{-1}$. Requiring that $\psi$ and $\partial_\mu \psi$ are continuous at $\mu_0$,
 one finds the following expressions for $\psi$:
\vspace{1cm}
\begin{eqnarray}
\psi_i&=&
\nonumber-\frac{S f^2}{2(\chi-1)}\Big(\chi^{-1}\cosh^2(\mu)\cos^2(\nu)\\
\label{psii}& & \quad\quad\quad\quad\quad+\chi\sinh^2(\mu)\sin^2(\nu)\Big),\\
\nonumber\psi_o&=&
-\frac{Sf^2}{4(\chi-1)^2}\Big[1+2(\mu-\mu_0)\\
\nonumber& & \quad\quad\quad\quad\quad+2(\chi-1)^2\sinh^2(\mu)\sin^2(\nu)\\
\label{psio}& &\quad\quad\quad\quad\quad+\frac{\chi-1}{\chi+1}\exp[-2(\mu-\mu_0)]\cos(2\nu)\Big],
\end{eqnarray}
where the subscripts $i$ and $o$ stand for inside and outside the core.

\begin{figure}
   \centering
   \includegraphics[width=1.0\linewidth]{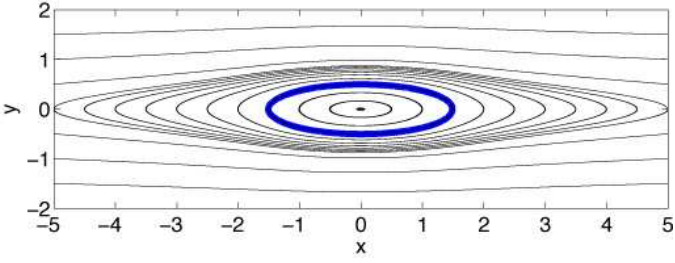}
   \caption{Vortex solution streamlines in a sheared flow with $\chi=3$. As expected, inside the vortex (delimited by the blue bold line) the streamlines are elliptical, with the same aspect ratio as the vortex itself. }
              \label{vortexsol}
\end{figure}

As expected, the inner solution reproduces the elliptical flow described in the previous subsection (Eqns.~\ref{v2Dx}---\ref{v2Dy}). In the outer solution, one recognises the background shear (3rd term), which dominates for large $\mu$. Moreover, this solution exhibits a linear term in $\mu$ corresponding to the long-range perturbation of the vortex. In cartesian coordinates, this linear dependance translates into a logarithmic tail for $\psi(x,y)$, showing that the vortex presence can be felt far from the core. As we will see in section \ref{Nonlinsim} below, this property leads to numerical artefacts when one tries to fit this solution in a finite-size box. For completeness, we show in Fig.~\ref{vortexsol} the streamlines obtained from solution (\ref{psii})---(\ref{psio}). As expected, the streamlines inside 
the core are ellipses of constant aspect ratio $\chi$. Outside the core, one still finds closed streamlines but with a much more elongated structure.

\subsection{Perturbation equations}
In the following, we concentrate on the evolution of perturbations in the vortex core, assuming the latter is infinite. This corresponds to a limit in which the perturbations are small compared to the typical horizontal size of the core.
To study the evolution of 3D perturbations in a 2D elliptical flow, we write the total velocity field as $\bm{u}=\bm{u}^0+\bm{v}$ where $\bm{v}$ is supposed to have an infinitely small amplitude. Following \cite{K80} and \cite{CC86}, we use a time explicit Fourier decomposition for the perturbation $\bm{v}=\bm{v}(t)\exp (i\bm{k}(t)\cdot \bm{x} )$.  Plugging this solution in (\ref{motiongeneral}) leads at first order to:
\begin{eqnarray}
\nonumber \dot{v}_i+ix_kv_i(\dot{k}_k + S k_j A_{jk} )&=&-ik_i \Pi-S v_jA_{ij}\\
\label{motionperturb}& & -2 \epsilon_{ijk}\Omega_j v_k\\
\label{divperturb} k_j v_j &=&0
\end{eqnarray}
where we have included the tidal term of (\ref{motiongeneral}) in $\Pi$. To satisfy (\ref{motionperturb}) for any $x_k$, one has to solve:
\begin{equation}
\dot{k}_k + S k_j A_{jk}=0,
\end{equation}
which leads to:
\begin{eqnarray}
\nonumber\bm{k}(t) &=&k_0 \Big(\sin(\theta) \cos\big[\phi(t)\big] \bm{e_x}\\
\nonumber & &-\chi\sin(\theta) \sin\big[\phi(t)\big] \bm{e_y}\\
\label{solution_k}& &+\cos(\theta) \bm{e_z} \Big)
\end{eqnarray}
where $k_0$, $\theta$ and $t_0$ are integration constants and $\phi(t)=S/(\chi-1)(t-t_0)$ is the turnover angle. 
Therefore, the perturbations will have a rotating wavevector with a turnover time equal to the vortex turnover time $T=2 \pi (\chi-1) /S$. Because of the incompressibility condition, $\theta=0$ will correspond to horizontal motion perturbations whereas $\theta=\pi/2$ will imply vertical ones. We also note that the rotating wavevector will involve smaller structures in the $y$ direction. According to this result, the horizontal aspect ratio of the perturbation wavevector is equal to that of the vortex ($\chi$).

We can then simplify (\ref{motionperturb}) eliminating the pressure, which leads to the final system
\begin{equation}
\label{eqfloquet}\frac{d v_i}{d\phi}=\Big[\Big(\frac{2k_ik_j}{k^2}-\delta_{ij}\Big)\bar{A}_{jm}+2\Big(\frac{k_ik_j}{k^2}-\delta_{ij}\Big)\bar{R}_{jm}\Big]v_m,
\end{equation}
where $\bar{\bm{A}}=(\chi-1)\bm{A}$ and $\bar{R}_{jm}=(\chi-1)\epsilon_{jlm}\Omega_l/S$.
Plugging solution (\ref{solution_k}) in this equation leads to a Floquet problem for $v$, as already pointed out by \cite{B86}. Interestingly, the Floquet problem doesn't depend on the norm of the wavevector $ k \equiv k_0,$ but just on its direction. Therefore, this problem is scale-independant, at least in the inviscid limit. The solution to this problem is known to be a superposition of Floquet modes written
\begin{equation}
\bm{v} = \exp(\gamma t) f[\phi(t)],
\end{equation}
where $f$ is periodic with period $2\pi$. To determine the Floquet exponents $\gamma$, we compute the eigenvalues $\exp(\gamma T)$ of the matrix $M(2\pi)$ where $M(\phi)$ satisfies the generalised equation
\begin{equation}
\label{eq_mat}\frac{d M_{in}}{d\phi}=\Big[\Big(\frac{2k_ik_j}{k^2}-\delta_{ij}\Big)\bar{A}_{jm}+2\Big(\frac{k_ik_j}{k^2}-\delta_{ij}\Big)\bar{R}_{jm}\Big]M_{mn},
\end{equation}
with the initial condition
\begin{equation}
\label{eq_mat_init} M_{ij}(0)=\delta_{ij}.
\end{equation}
A numerical approach is required to solve this problem in most cases. However, some limits can be understood using analytical approaches, as we will see in the next section.

\section{The elliptical instability}
\subsection{\label{horizontal_instability}Horizontal instability}
It is possible to derive an analytical criterion for the instability in the limit $\bm{k}=k_z\bm{e_z}$. In this limit, $v_z=0$ and (\ref{eqfloquet}) is reduced to
\begin{eqnarray}
\frac{dv_x}{d\phi} &=& \Big[-\chi + \frac{2\Omega}{S}(\chi-1)\Big]v_y,\\
\frac{dv_y}{d\phi} &=& \Big[ \frac{1}{\chi}-\frac{2\Omega}{S}(\chi-1) \Big]v_x.
\end{eqnarray}
This system describes horizontal epicyclic oscillations with frequency
\begin{equation}
\label{ling}
\kappa^2 = S^2\Big(R-\frac{\chi}{\chi-1}\Big)\Big(R-\frac{1}{\chi(\chi-1)}\Big),
\end{equation}
having defined the rotation number as $R=2\Omega/S$. In the limit $\chi \rightarrow \infty$ where the vortex is infinitely weak, we find the classical epicyclic frequency $\kappa^2=2\Omega(2\Omega-S)$, which is equal to the Keplerian frequency in discs. 
This epicyclic frequency is imaginary when
\begin{equation}
\frac{1}{2}+\frac{\sqrt{1+4/R}}{2}<\chi<\frac{R}{R-1}.\label{CRI}
\end{equation}
If a columnar vortex is in this regime, 
its horizontal layers will tend to drift exponentially
in the $(x,y)$ plane, leading to the destruction of the vertical
coherence of the structure. In accretion disc vortices,
this regime is found for $3/2<\chi<4$.
 
\subsection{A generalization}
We note that the above discussion may be generalized
to apply to a wider class of steady flows $(u_x^0 , u_y^0).$
The linearized equations of motion are
\begin{equation}
{\cal D} v_i +v_j\frac{\partial u^0_i}{\partial x_j}- 2\Omega\epsilon_{ijz}v_j= -\frac{\partial \Pi'}{\partial x_i}, \label{LINP} 
\end{equation}
where $\Pi'$ is the pressure perturbation.
The operator 
\begin{equation}{\cal D}\equiv \frac{\partial}{\partial t}
+u^0_j\frac{\partial}{\partial x_j}\end{equation}
denotes the convective derivative. 
We now assume perturbations are local in $z$ so that we may assume
that the $z$ dependence is through a factor $\exp(ik_z z),$ where $k_z$
is very large in the sense that the length scale $k_z^{-1}$
is smaller than any other in the problem.
From the incompressibility condition $\nabla\cdot{\bf v}=0,$
and the $z$ component of (\ref{LINP}) we conclude that $\Pi' = O(k_z^{-2}).$
Thus it may be dropped from the $x$ and $y$ components of (\ref{LINP})
which then give the pair
\begin{equation}
{\cal D}v_x +v_x \frac{\partial u_x^0}{\partial x}=
- v_y\left(\frac{\partial u^0_x}{\partial y}-2\Omega \right), \label{RLIN0}
\end{equation}
\begin{equation}
{\cal D}v_y +v_y\frac{\partial u^0_y}{\partial y} = 
-v_x\left(\frac{\partial u^0_y}{\partial x}+2\Omega \right).\label{RLIN}
\end{equation}
Replacing ${\cal D}$ by  ${d \over dt},$ with the understanding that
the time derivative is to be taken on a fixed streamline, we see
that in general the system (\ref{RLIN0})-(\ref{RLIN})
 leads to a second order ordinary differential
equation with periodic coefficients,  the period being the
time to circulate round the chosen streamline. But note that for the problem on hand the
coefficients  are constant
because the  unperturbed velocity is a linear function of the coordinates.
In more general cases one has a Floquet problem of the type described above
for every streamline indicating that unstable modes are indeed localized on streamlines.
In order to connect with (\ref{CRI}) we note that one can prove that
if everywhere  on a chosen streamline
\begin{equation} \left(\frac{\partial u^0_x}{\partial y}-2\Omega \right)
\left(\frac{\partial u^0_y}{\partial x}+2\Omega \right) >0,
\end{equation}
there will be an instability.
In fact for the problem on hand this is precisely equivalent to the above condition
that  $\kappa^2 <0$ which direcrly leads to (\ref{CRI}).
To show this we first note  that if 
\begin {equation} \alpha = \left(\frac{\partial u^0_x}{\partial y}-2\Omega \right)
\ \ {\rm and } \ \
\beta = \left(\frac{\partial u^0_y}{\partial x}+2\Omega \right) 
\end{equation}
then both $\alpha$ and $\beta$ must be either positive or negative.
Without loss of generality we assume both to be positive (the case when they
are both negative may be recovered by setting $v_x \rightarrow -v_x$).
Then from equations  (\ref{RLIN0})-(\ref{RLIN})  we may  obtain
\begin{equation} {d( v_xv_y)\over dt}    = -  \alpha v_y^2 - \beta v_x^2.
\label{RLIN3}\end{equation}
or setting $q_y=-v_y$
\begin{equation} {d( v_xq_y)\over dt}    =   \alpha q_y^2 + \beta v_x^2.
\label{RLIN4}\end{equation}
Thus if $\alpha_{min}$ and $\beta_{min}$ are the minimum values
of $\alpha$ and $\beta$ respectively, we have
\begin{equation} {d( v_xq_y)\over dt} \le  \alpha_{min} q_y^2 +
\beta_{min} v_x^2,
\label{RLIN5}\end{equation}
or equivalently
\begin{equation} {d( v_xq_y)\over dt} - 
2\sqrt{\alpha_{min}\beta_{min}}v_xq_y 
\le \left( \sqrt{\alpha_{min}} q_y -
\sqrt{\beta_{min}} v_x\right)^2.
\label{RLIN6}\end{equation}
Thus it follows that
$v_xq_y$ grows faster than exponentially with growth rate
$2\sqrt{\alpha_{min}\beta_{min}}$ 
which means that there is a 
Floquet exponent  corresponding  to exponential growth.
 
\subsection{Numerical results}
In the general case ($\theta >0$ and $\chi>1$), the Floquet problem (\ref{eqfloquet}) can't be solved analytically. We therefore use the following numerical approach.
We first evolve the $M$ matrix following Eq.~\ref{eq_mat} for a given $\chi$ and $\theta$ with a 4th order Runge-Kutta-Fehlberg algorithm, keeping the error down to $10^{-10}$. The eigenvalues are then computed from $M(2\pi)$ using the GNU scientific library routines. This procedure is repeated for 1000 values of $\chi$ and 1000 values of $\theta$ to get a 2D representation of the regions for which the instability is found.

\begin{figure}
   \centering
   \includegraphics[width=1.0\linewidth]{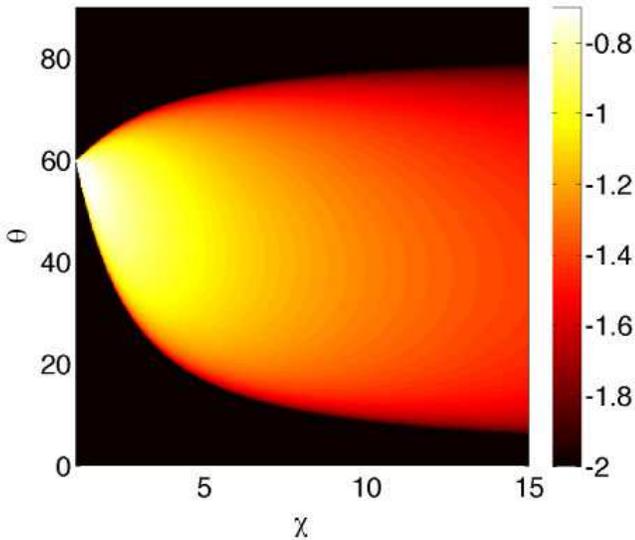}
   \caption{Growth rate (in shear time) of the 3D instability deduced numerically from the Floquet problem (\ref{eqfloquet}) in the non rotating case. This result is identical to the original elliptical instability described by \cite{B86} and allow us to test our numerical approach. The colour scale represent the logarithm of the growth rate. }
              \label{FloquetR0}
\end{figure}

To test our numerical results, we have first reproduced the 
results of \cite{B86} for a vortex in a non rotating sheared flow. 
One find in this case that vortices are 
unconditionally unstable to 
fully 3D disturbances (Fig.~\ref{FloquetR0}).
 Moreover, in the limit $\chi\rightarrow 1$, only modes with $\theta=60^\circ$ are
 found to be unstable, consistently with Bayly's findings.
 Finally, we note that the growth rate for the instability weakens as we elongate the vortex. This finding, although apparently opposite to the classical
 result for the elliptical instability, is
 due to the fact that our growth rate is measured in inverse shear time and not in 
inverse vortex turnover time.

\begin{figure}
   \centering
   \includegraphics[width=1.0\linewidth]{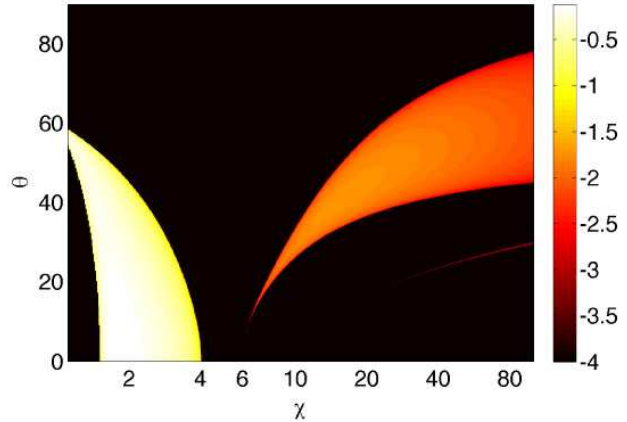}
   \caption{Growth rate of the 3D instability deduced numerically from the Floquet problem (\ref{eqfloquet}) in the Keplerian case for large $\chi$. One observes the vertical modes for $3/2<\chi<4$ and full 3D modes for $\chi<3/2$ and $\chi>6$ up to at least $\chi=100$. A weak secondary band is found under the primary band for large $\chi$.}
              \label{FloquetN0}
\end{figure}

We have used our numerical method to solve the Floquet problem in the Keplerian case (Fig.~\ref{FloquetN0}). In this case, we observe essentially a two bands structure. The first band (low-$\chi$ band) is located between $\chi=1$ and $\chi=4$. It is associated with a short growth time (typically of the order of one shear time), 
and it includes the horizontal instability described in
 section \ref{horizontal_instability} for $\theta=0$. 
In the limit $\chi\rightarrow 1$ (infinitely strong vortex), 
we find the same result as in the non rotating case
 (instability for $\theta=60^\circ$), which was to be expected since 
the vortex turnover time is much smaller than the rotation time in this limit.
 The second band (high-$\chi$ band) is found for $\chi>6$ and is much weaker
 since it involves growth rate of the order of $10^{-2} S$.
 This band gets wider and weaker as we go to large aspect ratios (and weak vortices),
 with a growth rate maximum found for $\chi\simeq 11.$
 We have tried to get a larger resolution in the region $\chi\sim 5.5$---6 and it seems that this band reaches $\theta=0^\circ$ for $\chi\sim 5.9$, but the growth rates involved are very small. Note that another narrower band is observed in the regime of high $\chi$ and smaller $\theta$ in Fig.~\ref{FloquetN0}. As we will see in the following, this band is due to higher order resonance and is therefore weaker than the bands described previously. 

\begin{figure}
   \centering
   \includegraphics[width=1.0\linewidth]{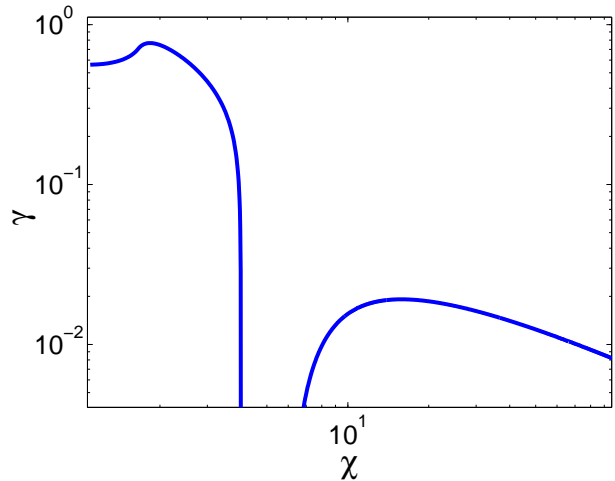}
   \caption{Maximum growth rate in the Keplerian case  (in shear time). We remark the low-$\chi$ band with large growth rates ($\chi<4$), the high-$\chi$ band with smaller growth rates ($\chi>6$) and a stable region in between. The growth rate decays for large $\chi$ when the vortex gets weaker, as expected.}
              \label{N0FloquetMax}
\end{figure}

To get a simpler representation of the elliptical instability in the Keplerian case, 
we have plotted the maximum growth rate as a function of $\chi$
 in Fig.~\ref{N0FloquetMax}. 
We observe the same basic features as on the colormaps with 2 bands. 
We also observe that the high-$\chi$ band decays as $\chi$ gets larger.
 This property is to be expected since the limit $\chi\rightarrow \infty$ corresponds to a pure shear flow, which is known to be linearly stable. 
However, for a finite (but large) $\chi$, 
vortices are always unstable, contrary to claims
in the literature \citep{Li07}.

Therefore, it seems that no elliptical instability is found for $4<\chi<5.9$. 
However, we recall that up to now  we have only considered perturbations interior to
the vortex core. But as we will see later, instabilities may also appear \emph{outside} the vortex core.  

\subsection{Physical interpretation\label{physinterp}}

It may be noted that the system (\ref{eqfloquet}) can be written as a single Hill's equation \citep[see for example][]{W90}. In this case, one finds that the periodic excitation frequency is twice the vortex frequency $S/(\chi-1)$, since the
time-dependent wave numbers $k_x$ and $k_y$ always appear quadratically. In the limit $\theta \rightarrow 0$ for which the excitation is small, one would expect a series of instability bands.  By analogy with Mathieu's equation, one should find such a
band when the local epicyclic frequency $\kappa$ is an harmonic of the vortex frequency:
\begin{equation}
\label{rescondition}
\kappa=n\frac{S}{\chi-1},
\end{equation}
with 
$n \  \epsilon  \ N$.
 In the Keplerian case, one would therefore expect instability bands arising from $\theta=0$ for $\chi_{1}\simeq 4.65$, $\chi_{2}\simeq 5.89$, $\chi_{3}\simeq 7.28$\dots\ However, according to Fig.~\ref{FloquetN0}, the first unstable band $\chi=4.65$ $(n=1)$ doesn't appear. This peculiar property can be understood writing the epicyclic frequency as
\begin{equation}
\kappa^2=S^2\Big[R\Gamma+\frac{1}{(\chi-1)^2}\Big],
\end{equation}
where $\Gamma=\omega_t/S+R$ is the absolute vorticity of the background vortex. According to this expression, the $n=1$ resonance condition is equivalent to the constraint $\Gamma=0$. Interestingly, the absolute vorticity also appears explicitly in the linearized vertical vorticity equation:
\begin{equation}
\partial_t \omega_z+\bm{u_0 \cdot \nabla} \omega_z = S \Gamma \partial_z v_z
\end{equation}
where $\omega_z=\partial_x v_y-\partial_y v_x$ is the vertical vorticity of the perturbation. According to this equation, $\omega_z$ is a conserved quantity if the background absolute vorticity $\Gamma$ is zero. We conclude from this argument that the $n=1$ resonance condition cannot lead to an instability, since $\omega_z$ could not be conserved in that case. Therefore, the first unstable band appears for $n=2$ ($\chi\simeq 5.89$), which is consistent with our numerical result. As stated before, higher order bands ($n>2$) are also observed in the computation (Fig.~\ref{FloquetN0}), but they are much weaker, and one can't check the origin of these bands on the $\theta=0$ axis with enough precision to compare with eq.~(\ref{rescondition}). 

\section{Stratified case}
\subsection{Model and Equations}
\cite{BM05} showed using anelastic numerical simulations that off-midplane vortices were able to survive for several hundred orbits whereas midplane vortices were destroyed rapidly. This suggests that stratification might be a way to stabilise vortices and suppress the elliptical instability. To study this problem in a simple configuration, we have considered the case of a 3D vortex centered at $z=z_0$, i.e. above the midplane. We then follow the evolution of perturbations with $k\gg z_0^{-1}$ inside the vortex. In first approximation, the evolution of these perturbations can be described using the Boussinesq approximation which can be written
\begin{eqnarray}
\label{motiongeneralB} \partial_t \bm{u}+\bm{\nabla\cdot} (\bm{u\otimes u})&=&-\bm{\nabla} \Pi
-2\bm{\Omega \times u}+2\Omega S y \bm{e_y}\\
\nonumber & &+N\zeta\bm{e_z} ,\\
\partial_t \zeta+\bm{\nabla\cdot} (\bm{u} \zeta)&=&-N u_z,\\
\label{divvB} \bm{\nabla \cdot u}&=&0,
\end{eqnarray}
where $\zeta$ is the potential temperature and $N$ is the Brunt-V\"ais\"al\"a frequency. In the following, we will assume a stable vertical stratification, i.e. a real Brunt-V\"ais\"al\"a frequency. The vortex solution found previously is still valid in the Boussinesq approximation. Therefore, we can follow the same procedure as the one used in the non stratified case to get
\begin{eqnarray}
\label{eqfloquetB}\frac{d v_i}{d\phi}&=&\Big[\Big(\frac{2k_ik_j}{k^2}-\delta_{ij}\Big)\bar{A}_{jm}+2\Big(\frac{k_ik_j}{k^2}-\delta_{ij}\Big)\bar{R}_{jm}\Big]v_m\\
\nonumber & &-\frac{N(\chi-1)}{S}\Big(\frac{k_i k_j}{k^2}-\delta_{ij}\Big)\zeta \delta_{jz},\\
\frac{d \zeta}{d\phi}&=&-\frac{N(\chi-1)}{S}v_z.
\end{eqnarray}
These equations associated with the solution (\ref{solution_k}) lead to a stratified Floquet problem.
As previously, this problem can be solved by diagonalizing a $4\times4$ evolution matrix $M$, following the same procedure as in the non stratified case. The numerical method used to solve these equations is the same as in the previous section, except we have added an extra dimension in the solver to handle the evolution of the potential temperature. 
\subsection{Results}
\begin{figure}
   \centering
   \includegraphics[width=1.0\linewidth]{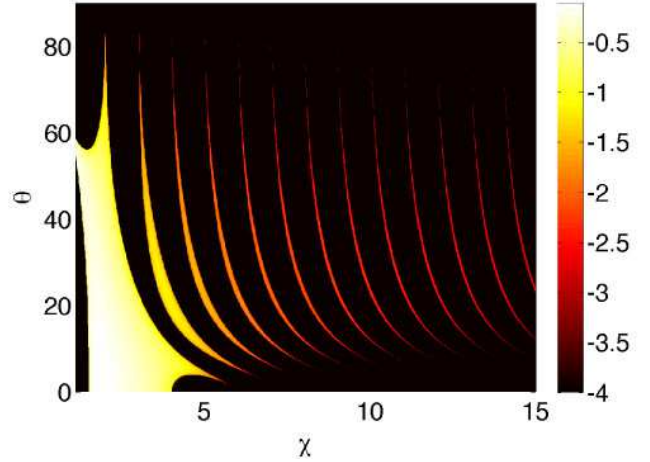}
   \caption{Growth rate of the 3D instability deduced numerically from the Floquet problem (\ref{eqfloquetB}) in the Keplerian case with $N/S=1.0$. One observe once again the vertical modes for $3/2<\chi<4$ and many weak bands involving fully 3D perturbations. The colour scale represent the logarithm of the growth rate.}
              \label{FloquetN1}
\end{figure}

We plot the growth rate of the instability as a function of $\chi$ and $\theta$ for a moderately stratified vortex $N=S$ in Fig.~\ref{FloquetN1}. We note that the influence of the stratification is strong in the domain in which the vortex is weak ($\chi>5$), and significant differences are observed when comparing with Fig.~\ref{FloquetN0}. First, we remark that the stable region observed in the non stratified case ($4<\chi<5.9$) is not present in the $N=S$ stratified case. This result is due to the presence of high frequency buoyancy modes which can match the turnover frequency when the inertial modes cannot. Moreover, the high-$\chi$ band has been replaced by a series of weaker bands. These bands can be understood as reminiscences of the resonance condition (\ref{rescondition}), 
since in the limit $\theta \rightarrow 0$ the effect of  stratification disappears.

\begin{figure}
   \centering
   \includegraphics[width=1.0\linewidth]{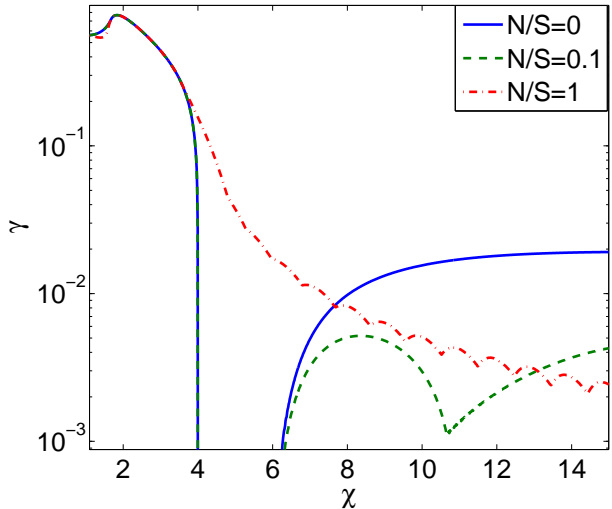}
   \caption{Maximum growth rate in the Keplerian case  (in shear time) for various stratification frequencies. As the stratification gets stronger, the growth rate decreases for weak vortices ($\chi>6$). We also note that for a strong enough stratification, the region without instability ($4<\chi<5.9$) disappears. Finally, the horizontal instability described in the previous section ($3/2<\chi<4$, $\theta=0$) is not affected by the stratification, as expected. }
              \label{FloquetN1Max}
\end{figure}

When we compute the maximum growth rate as a function of $\chi$ for various 
stratification frequencies (Fig.~\ref{FloquetN1Max}), 
we note that the growth rate at large $\chi$ decreases significantly with stronger stratification. 
In this sense, the stratification tends to weaken the elliptical instability, but does not suppress it.


\section{Non-linear simulations}\label{Nonlinsim}
\subsection{Resolving the elliptical instability}
As pointed out in the previous section, the elliptical instability is always present for highly elongated vortices ($\chi > 6$), 
although it is quite weak.
We have also found that no elliptical instability was observed for $4<\chi <5.9$ when stratification was omitted.
To check these results, one has to perform non-linear simulations in which this instability is resolved. 
A condition to get this instability in a numerical simulation may be found assuming we have a 2D vortex embedded in a disc with a height $H$, the vortex core being larger than $H$. In the limit $\chi\gg 6$, 
we find the elliptical instability for $\theta\sim \pi/3$ (Fig.~\ref{FloquetN0}). 
According to (\ref{solution_k}), unstable modes will have in this case $k_x^\mathrm{max}\sim k_z$
and $k_y^\mathrm{max}\sim \chi k_z$. Moreover, $k_z>2\pi/H$ since 3D perturbations have to fit
vertically inside the disc. If we assume we need $n$ points to
resolve one wavelength ($n>2$), 
then the longest wavelength modes unstable to the elliptical instability will be resolved if the $x$ resolution
is $n$ points per scale height and the $y$ resolution is $n\chi$ points per scale height. 
This last condition is not often satisfied in 3D global simulations,
and may explain why long-lived and stable vortices are sometimes observed.
Note that the value $n$ may be 
quite high when using low order finite difference methods (one might expect $n\sim 10$) since one wants the 
(numerical) dissipation rate for these wavelengths to be smaller than the growth rate, 
which is itself small compared to the shear frequency. 
Therefore, \emph{very accurate} (e.g. spectral) numerical methods are preferable to study this kind of weak instability.

To check for resolution artefacts, the non stratified incompressible simulations and 
the compressible simulations carried out with NIRVANA below
have been computed with two different resolutions (the ``high'' resolution being twice the ``low'' resolution in each direction). No significant difference 
was found between high resolution and low resolution runs. In particular, the same growth rates were obtained,
with the same localisation properties for the instabilities. However, in low resolution runs carried 
out with NIRVANA that were seeded with truncation errors, the unstable scales were initially a factor $\sim 2$ larger, 
but with the growth rates and later
nonlinear behaviour being very similar. This can be understood as a consequence of the
scale free property of the linear instability that holds once scales are sufficiently small
but still larger than the dissipation scale, here expected to be the grid scale.
Noting that conclusions would be unaltered if the low resolution runs were adopted, only high resolution results
will be presented in this section.

\subsection{\label{nrsol}Numerical solution for an isolated vortex}

\begin{figure}
   \centering
   \includegraphics[width=1.0\linewidth]{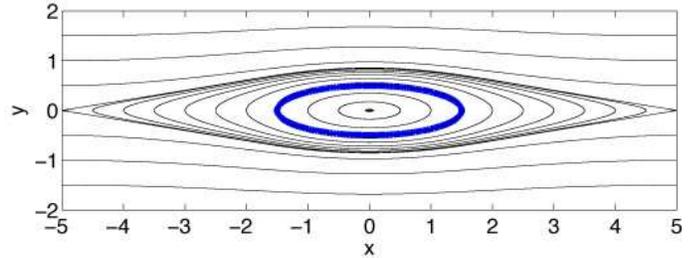}
   \caption{Vortex streamlines in a sheared flow with $\chi=3$. This solution is computed with an FFT method, assuming periodic boundary conditions in $x$ and $y$. This solution and the infinite solution (Fig.~\ref{vortexsol}) are extremely similar, except around $(x=\pm 5,y=0)$ where an X point is observed.}
              \label{vortexsolperiodic}
\end{figure}

To check the existence of the elliptical instability in accretion disc vortices, one needs a solution for an isolated vortex. In section \ref{model_section}, we have derived an exact solution for a vortex in an infinite box. Numerically however, one has to work in a finite size domain, using essentially the shearing-sheet boundary conditions \citep{HGB95}. As pointed out previously, the infinite solution has a slowly decaying tail which can lead to artefacts when one starts with this solution in a finite-size shearing box. When one tries this procedure, the 2D solution rapidly develops instabilities near the boundaries, especially along the $x$ axis, leading to large amplitude oscillations. To prevent this, we have chosen to solve (\ref{poisson}) using fast Fourier transforms (FFT), 
which are a natural choice for our boundary conditions. A solution obtained by this method in a $(10\times 4)$ box is shown in Fig.~\ref{vortexsolperiodic}. The solution found by this procedure is very similar to the infinite solution, except near $(x=\pm 5,y=0)$ where an $X$ point is observed. This difference probably explains the unsteadiness observed when one uses the infinite solution numerically. We find that periodic solutions derived using the FFT procedure are better suited for numerical simulations. Although these solutions are not \emph{exactly} steady (mainly because of time varying boundary conditions), they keep the same shape and the same amplitude for several turnover times when one uses high precision numerical methods, which is enough for our purpose.

As pointed out previously, this vortex model is not a proper infinite elliptical flow, 
and the approximation used in our linear analysis may not be valid. 
Nevertheless, the core of these elliptical vortices is made of elliptical streamlines. 
Since our linear analysis is essentially a local analysis of stability on  individual streamlines, one expects 
our results to apply to the core of such vortices. 
Note however that other parametric instabilities may appear \emph{outside} of the vortex core 
 because closed streamlines still exist in this region (see Fig.~\ref{vortexsolperiodic}).

We have computed the evolution of several isolated elliptical vortices in a shearing box with dimensions 
$L_x\times L_y\times L_z$ using an incompressible spectral method \citep[see][for a complete description 
of the numerical method]{LL05}. 
We have used $(N_x\times N_y\times N_z)=(256\times 512\times 64)$ points and an eighth order
 hyperviscosity instead of the classical viscosity to prevent the dissipation of the weaker unstable modes. 
The vortices are initialised in the center of the box setting a vorticity patch with dimension ($\lambda_x,\lambda_y$).
 We always set $\lambda_y=1$, $L_y=4$ and $L_z=2$, $\lambda_x$ and $L_x$ being adjusted according to the vortex aspect-ratio $\chi$. For $\chi=3$, we set $L_x=10$ whereas for $\chi=5.5$ and $\chi=11$ we set $L_x=20$. For each run, we initialise the velocity field solving (\ref{poisson}) in Fourier space, and we add a small ($10^{-7}$) 3D white noise to the 2D solution. 

\begin{figure}
   \centering
   \includegraphics[width=0.9 \linewidth]{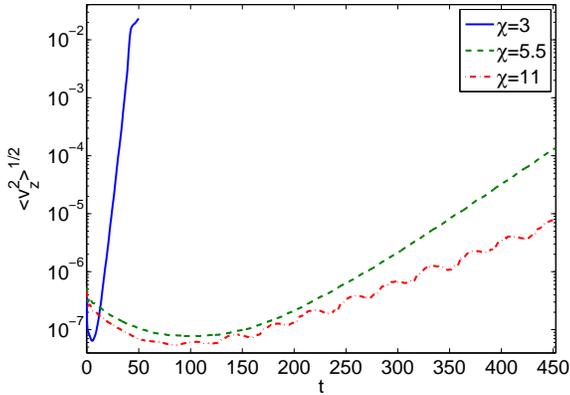}
   \caption{Time evolution of $\sqrt{\langle v_z^2\rangle}$ for several isolated vortices in shearing box simulations with no stratification. We find an exponential growth in agreement with the linear prediction for $\chi=3$ and $\chi=11$. We also find an instability for $\chi=5.5$ although the elliptical streamlines are stable in our linear analysis.}
              \label{vzenerg}
\end{figure}

\begin{figure*}
   \centering
   \includegraphics[width=0.40 \linewidth]{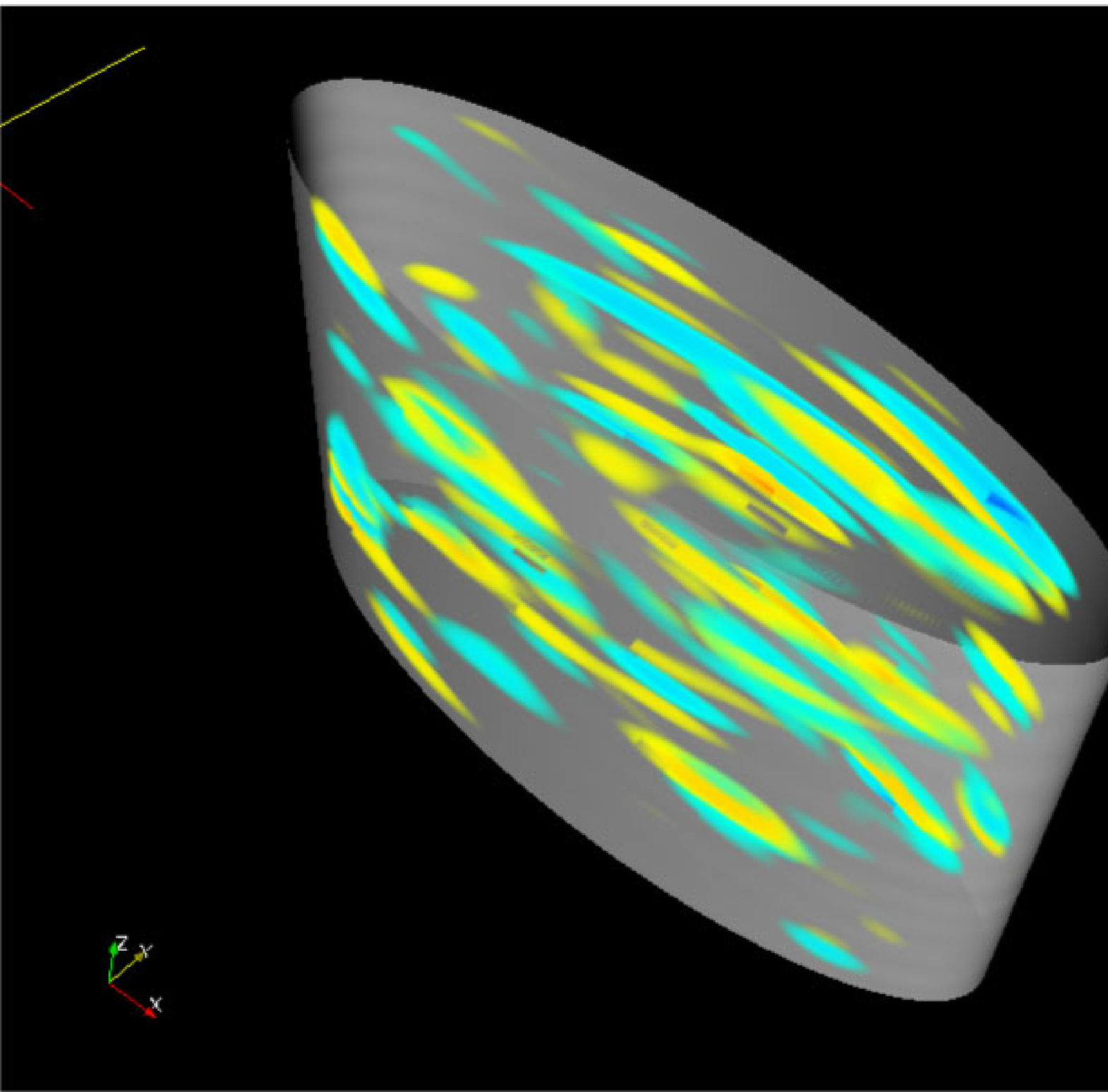}\quad
   \includegraphics[width=0.40 \linewidth]{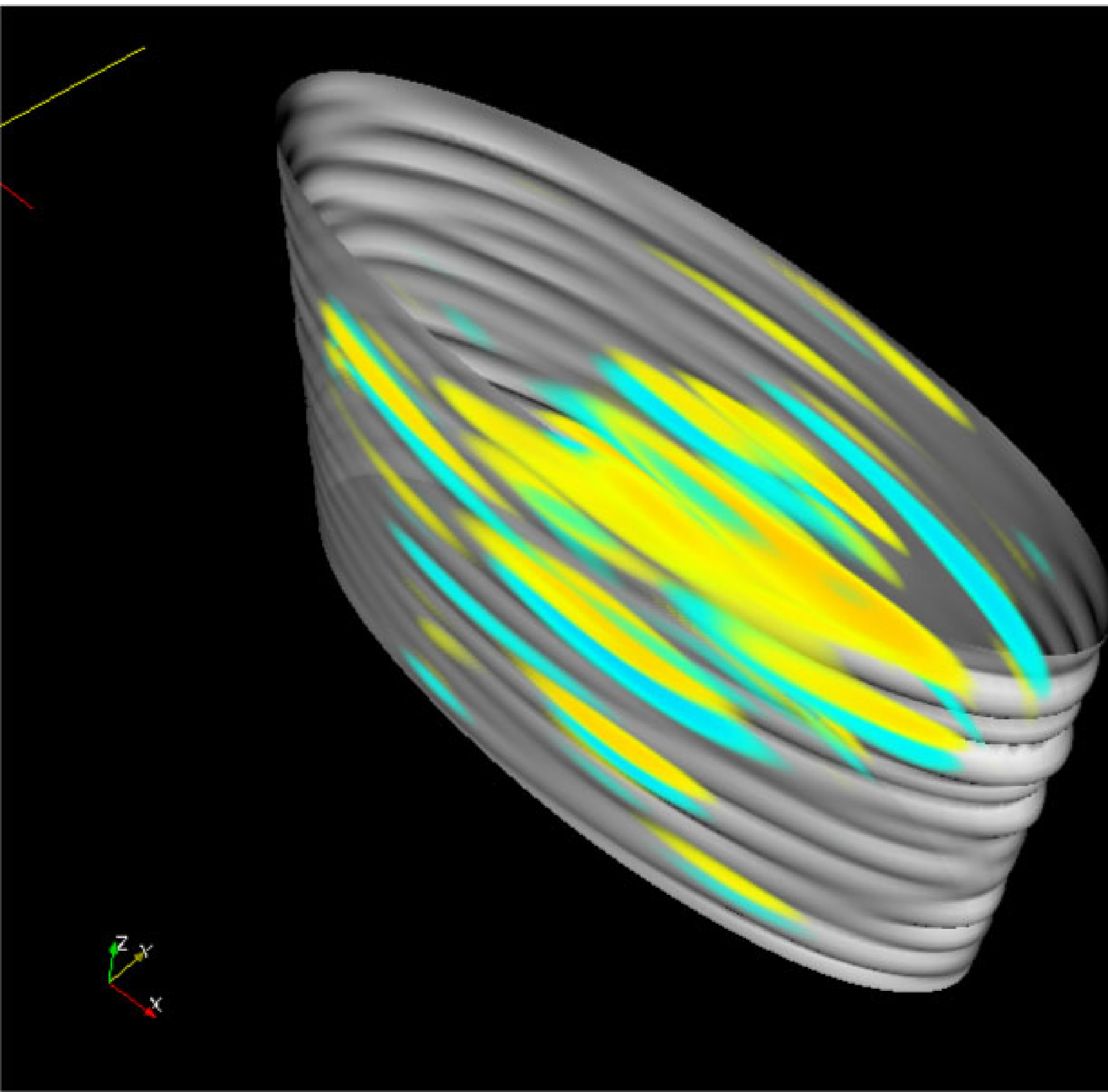}\\ \quad \\
   \includegraphics[width=0.40 \linewidth]{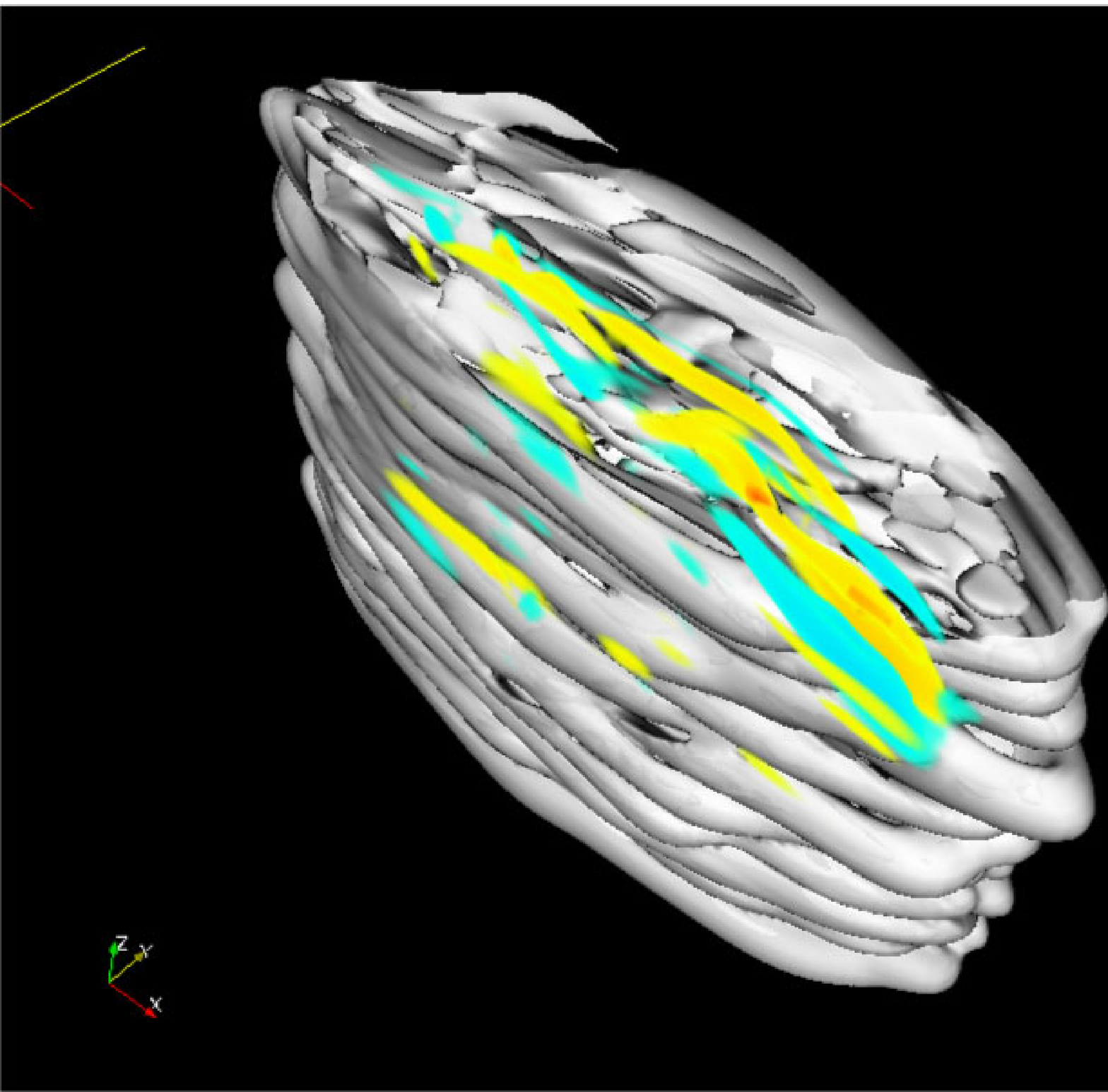}\quad
   \includegraphics[width=0.40 \linewidth]{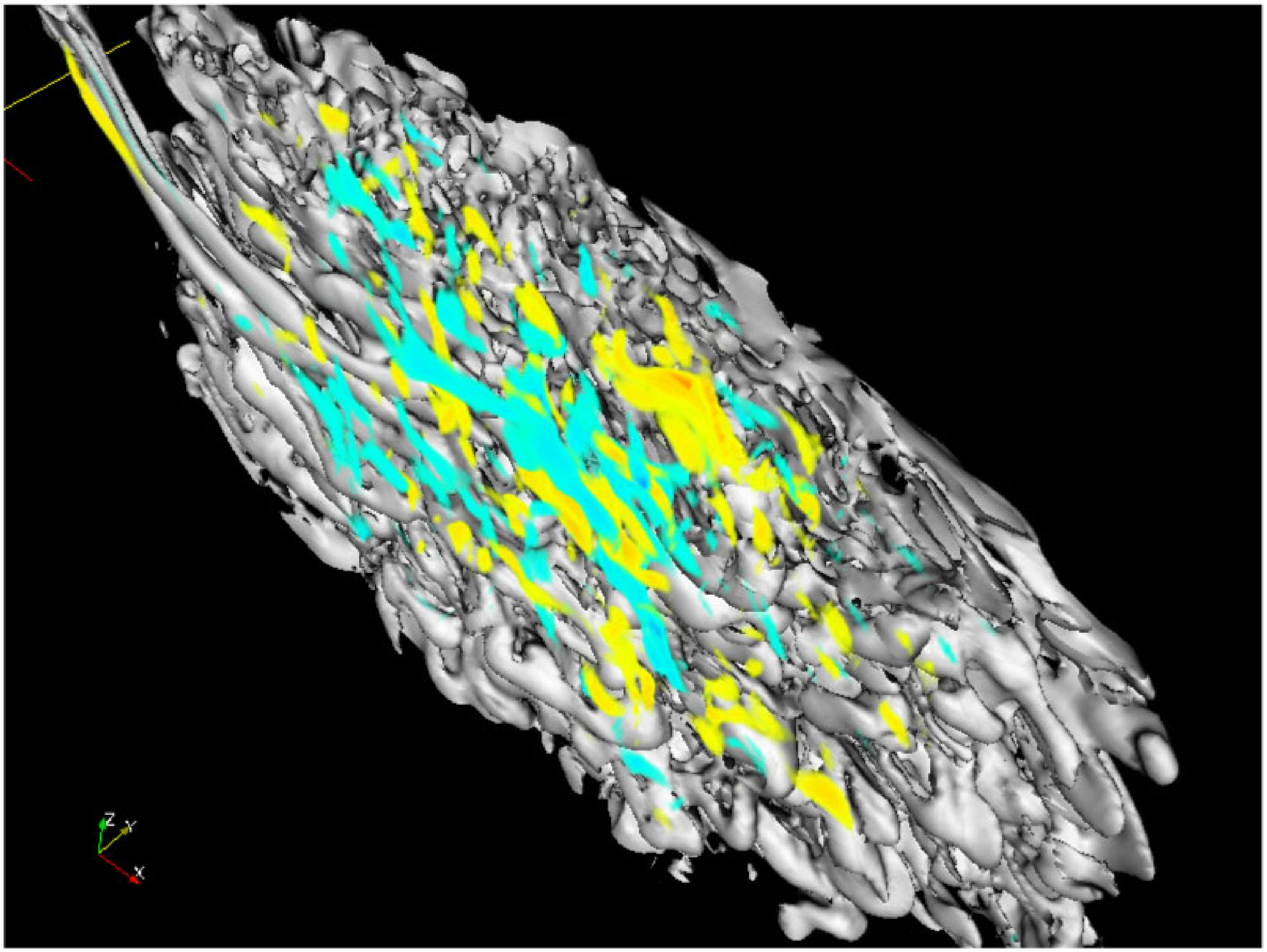}
   
   \caption{3D snapshots showing the evolution of a $\chi=3$ vortex at $t=26$ (top left), $t=36$ (top right), $t=41$ (bottom left) and $t=47$ (bottom right). We have represented in transparent grey an isocontour of vorticity delimiting the vortex core, and in colour the volume rendered vertical velocity normalised. We find that the instability is localized inside the vortex and is dominated by small scale vertical wave-numbers ($\theta\sim 0$), as expected. Moreover, the vortex structure is destroyed when the instability saturates, i.e. for $t\simeq 40$. }
              \label{destruction}
\end{figure*}

\begin{figure*}
   \centering
   \includegraphics[width=0.6 \linewidth]{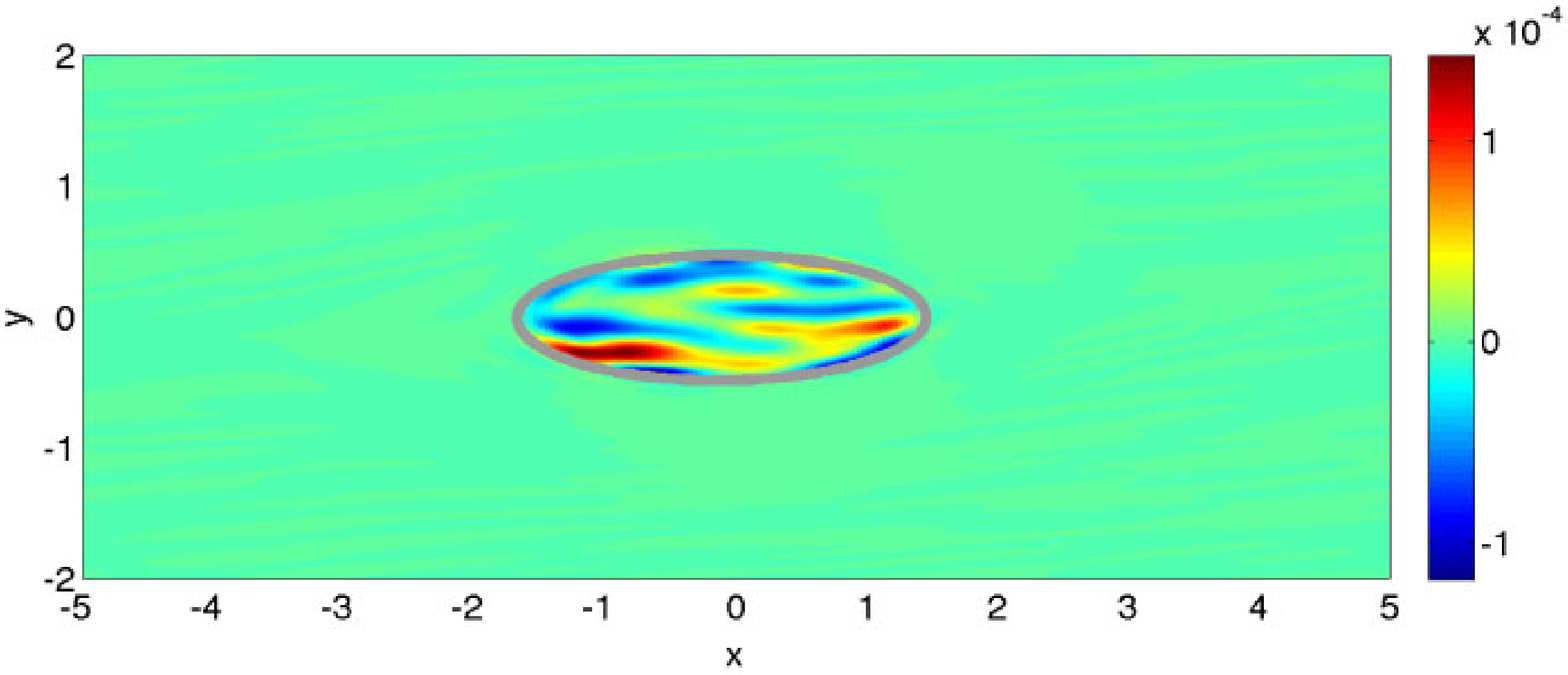}
   \includegraphics[width=0.99 \linewidth]{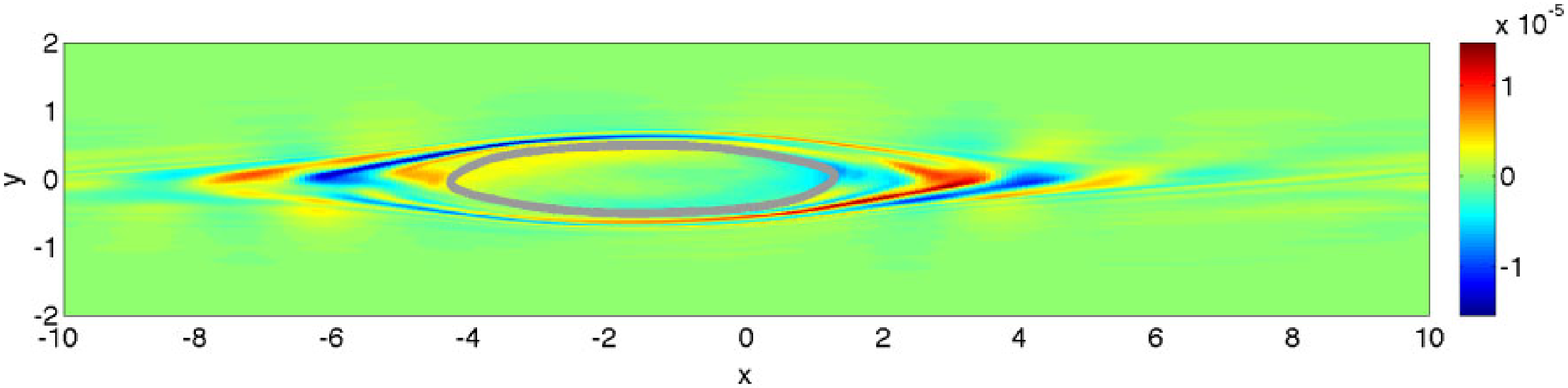}
   \includegraphics[width=0.99 \linewidth]{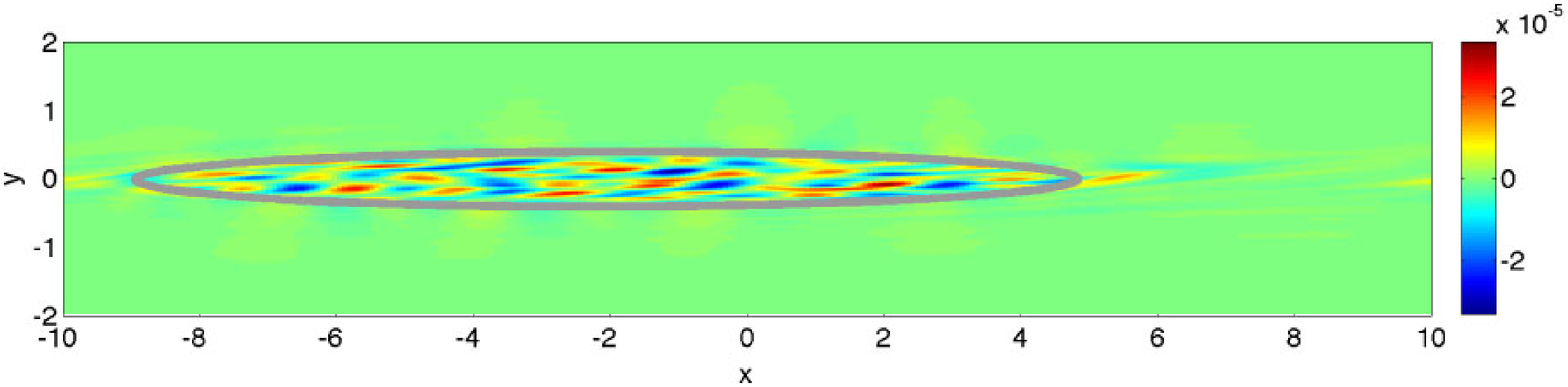}
   
   \caption{2D snapshots of the vertical velocity field in the $(x,y)$ plane for $\chi=3$ (top), $\chi=5.5$ (middle) and $\chi=11$ (bottom). The vortex core is  delimited by a thick grey line. We find that 3D perturbations are localized inside the vortex core for $\chi=3$ and $\chi=11$ whereas they are found outside of the vortex core for $\chi=5.5$. }
              \label{vzloc}
\end{figure*}

\subsection{Non stratified case}

To follow the evolution of 3d motions, we have computed $\sqrt{\langle v_z^2\rangle}$, where $\langle \rangle$ denotes a volume average procedure. Time histories of this quantity are given in Fig.~\ref{vzenerg} for $\chi=3$, $\chi=5.5$ and $\chi=11$. As expected from our linear analysis, we find an exponential growth for both $\chi=3$ and $\chi=11$. Since the growth rate for $\chi=3$ is very fast, the instability saturates at $t\simeq 40$ and the vortex structure is destroyed, as shown in Fig.~\ref{destruction}. In the exponential regime, we find a growth rate $\gamma=0.37$ for $\chi=3$ and $\gamma=0.016$ for $\chi=11$, in agreement with theoretical values ($\gamma_3=0.44$ ; $\gamma_{11}=0.017$). We also find an instability for $\chi=5.5$ with $\gamma_{5.5}=0.055$ which was not expected in our linear analysis ($\chi=5.5$ elliptical streamlines are stable according to Fig.~\ref{N0FloquetMax}). 

To check the localisation of the instability, we present $(x,y)$ slices of the vertical velocity for $\chi=3$, $\chi=5.5$ and $\chi=11$ in Fig.~\ref{vzloc}. We find that unstable modes are localised inside the vortex for $\chi=3$ and $\chi=11$, as expected from our linear analysis. However, for $\chi=5.5$, the instability appears outside of the vortex core, in a region in which one find closed non-elliptical streamlines (see Fig.~\ref{vortexsol}). In this particular case, no vertical motions are found in the vortex core, which is consistent with our analysis.

The existence of this outer instability can be understood 
quite simply following the physical description used for
 the elliptical instability (section \ref{physinterp}).
 We know that these instabilities arise when the epicyclic frequency
 is an harmonic of a closed streamline frequency multiplied by $2\pi$
(the streamline frequency being the inverse of the time needed by a fluid particle 
to cover the whole streamline and get back to its starting point).
Moreover, as one moves away from the vortex core,
the streamline frequency tends to 0 since the flow tends
towards a pure shear flow, whereas the epicyclic frequency
 tends to the Keplerian frequency.
 Therefore, one will always be able to find a
 resonance outside of the vortex core,
 and consequently a parametric instability.

\section{Effects of compressibility}

\begin{figure}
   \begin{center}
   \includegraphics[width=0.6 \linewidth,angle = 270]{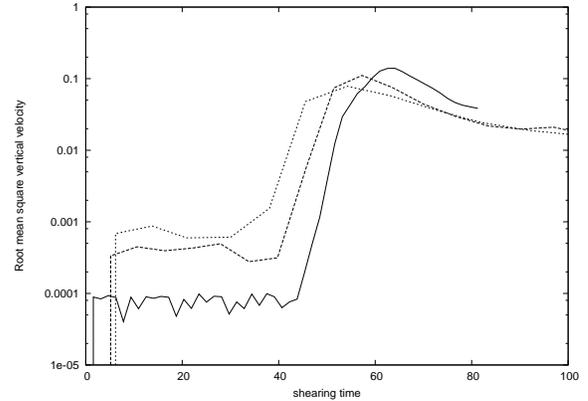}
   \caption{The root mean square vertical velocity obtained from the
   compressible simulations is plotted as a function of time
    for $\chi=2$.
   The cases shown are $v_s/c_s=1.3$ (dotted line), 
 $v_s/c_s=0.65$ (dashed line),  and  $v_s/c_s = 0.13$
   (full line).  Initial values increase with $v_s/c$ because
   the initially imposed state variables are  further from being in 
a stationary state.}
    \hfill   \label{jcbp1}
\end{center}
\end{figure}

\begin{figure*}
   \begin{center}
   \includegraphics[width=1.0  \linewidth]{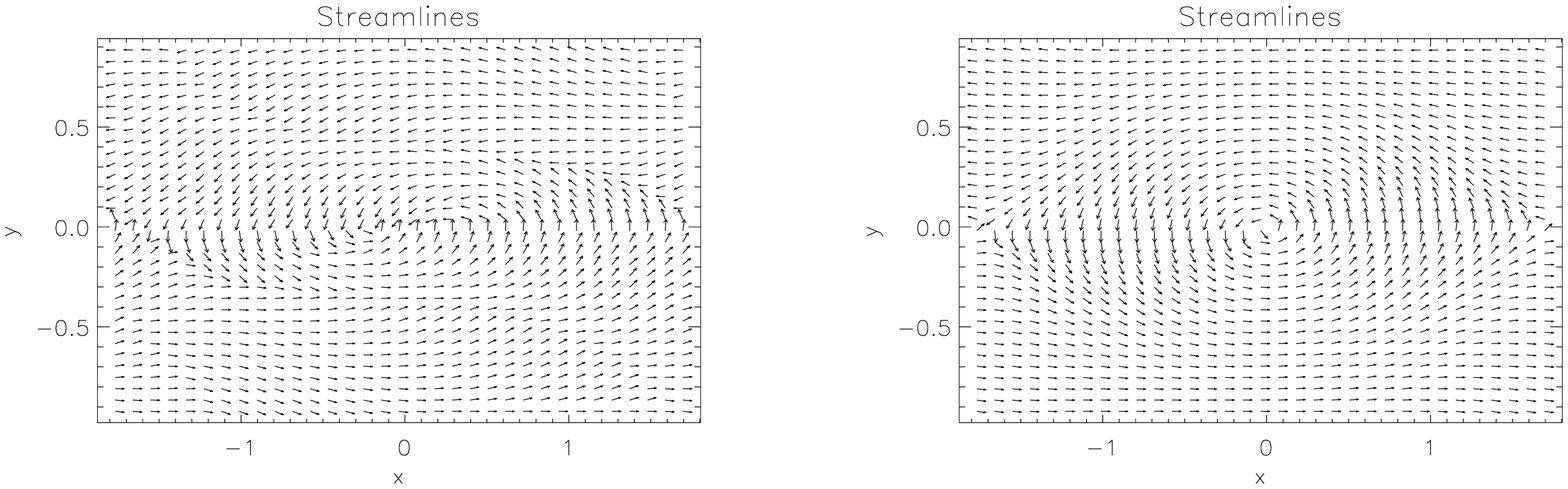}
   \caption{ Streamlines in the $(x,y)$ plane
    for $\chi = 2$
   obtained from the compressible simulations  shown 
   after the onset of linear instability. These are with 
$v_s/c_s=1.3$ at $t= 37$  (left), and  $v_s/c_s = 0.13$ at $t=47$ (right). 
   The initially imposed flow  is well preserved
    when $v_s/c_s = 0.13.$ When $v_s/c_s = 1.3$ the flow
    shows greater time variability, with the vortex being sqeezed
    in the radial direction. Some trailing  features are established outside
    the initial vortex core.} 
   \hfill  \label{jcbp2}
 \end{center}
 \end{figure*}
 
 \begin{figure*}
     \begin{center}
   \includegraphics[width=0.8 \linewidth]{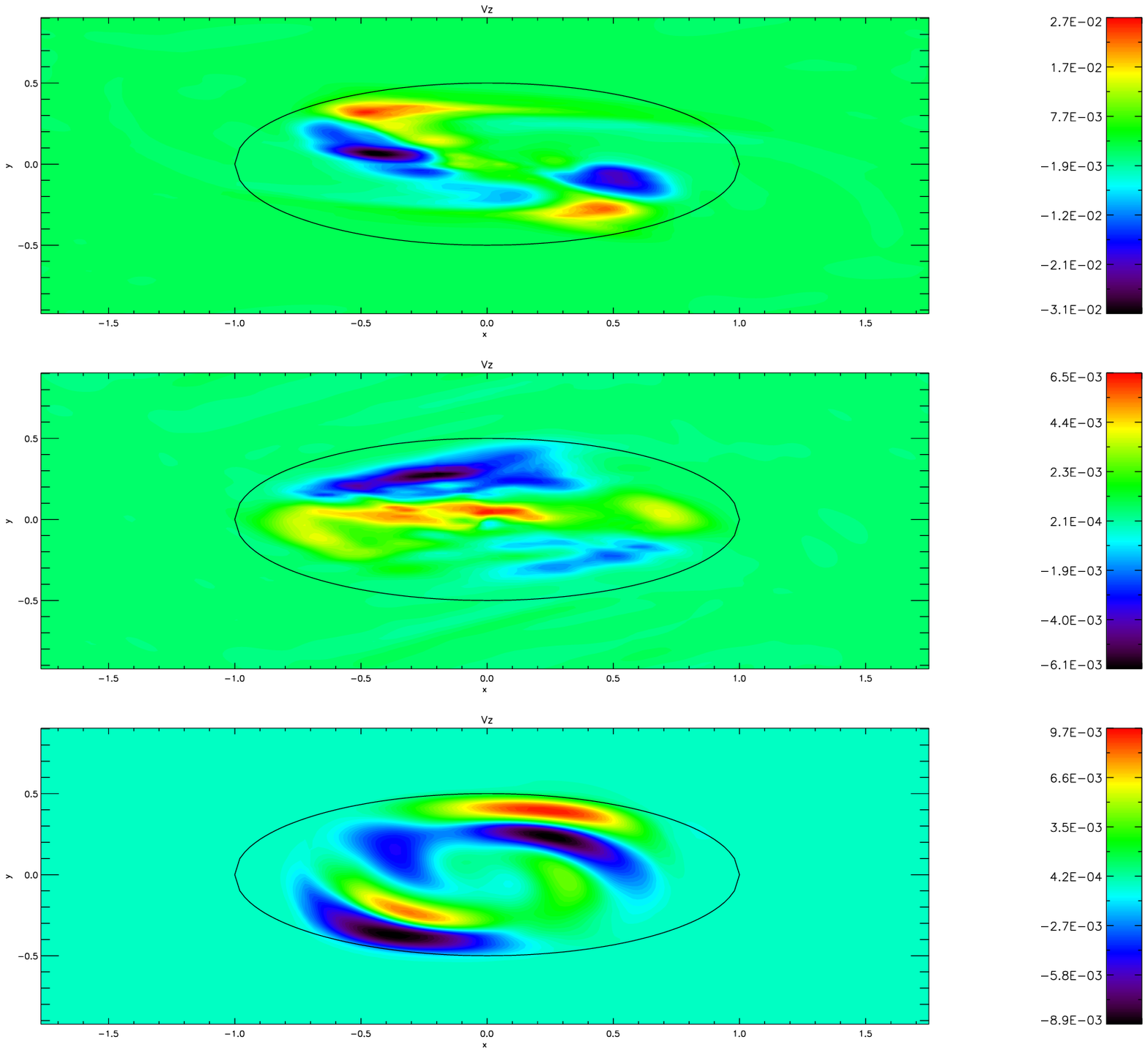}
   \caption{  2D snapshots of the vertical velocity field in the $(x,y)$ plane
    for $\chi=2$
   obtained from the compressible simulations shown after
 the onset of linear instability.
 These are with $v_s/c_s=1.3$ at $t= 37$
  (top), $v_s/c_s = 0.65$ at $t=42$ (middle) and $v_s/c_s = 0.13$ at $t=47$.
 The elliptical boundary of the vorticity patch is overplotted.
 We find that the 3D perturbations are localized inside this
 boundary as in the incompressible case.}
   \hfill      \label{jcbp3}
\end{center}
\end{figure*}

We have also investigated the effects of compressibility.
To do this we performed three dimensional hydrodynamic simulations
using NIRVANA \citep{ZY97}.
NIRVANA  has been used frequently in the past to study
various problems involving MHD turbulence in the shearing box
\citep{FP06,PNS04}. Because of its diffusive
character, it is best suited to vortices with small $\chi$ that are not
too elongated. For this reason we limit consideration to $\chi=2.$
We consider isothermal shearing boxes 
with vertical gravity. In units such that
the  full radial  width of the elliptical vorticity patch is unity,
the box dimensions were $(L_x= 3.46  ,L_y= 1.73  ,L_z= 2.31 ).$
The numbers of grid points used were  $(N_x= 196 ,N_y= 98 ,N_z= 128 ).$
A vertical gravitational acceleration $-\Omega^2 z$ was applied
over 80\% of the vertical computational domain, being set to zero
in two domains extending from the vertical boundaries, each being of an
extent equal to 10\% of the whole vertical domain. The boundary conditions
of periodicity in shearing coordinates could then be applied.
We comment that the use of periodic boundary conditions in the vertical direction,
with small buffer zones near the boundaries that allow these conditions to be applied 
consistently,
has been found to be effective in vertically stratified shearing box simulations
of MHD turbulence \citep[eg.][]{SHGB96}.
This approach allows for regular treatment of the boundary without significant
artefacts being produced in the mid-plane regions of the  flow, as would be expected for
local instabilities of the type considered here. To test this as well as  the effects 
of expanding the computational
domain in general, we have performed a test simulation with the compuational domain doubled in size
in each direction for the case with the largest value of $v_s/c_s$ considered below.
The dimensions of the elliptical vorticity patch were not changed.
The growth rate and nonlinear development of the  instability were indeed essentially the same
in the mid-plane regions, which because of  higher inertia carried most of the energy,
as in the smaller domain simulations presented below. The instability in the larger domain case 
 readily spreads
to the more extensive regions outside the original vorticity patch. 

We considered three values of the isothermal sound speed $c_s.$
Adopting the shearing time as the unit of time, these were such that
$v_s/c_s$ were $1.3, 0.65,$ and $0.13.$ Here the velocity
$v_s,$ being unity in our dimensionless units is the maximum
velocity difference across the vorticity patch due to the
background shear.

The incompressible  two dimensional flow field calculated in
section \ref{nrsol} was applied
on horizontal planes. As above this was calculated by solving the vorticity
equation with the  imposition of  periodic boundary conditions.
To make this possible, one must subtract out the mean vorticity in the patch
so that the net surface integral over the $(x,y)$ plane is zero, which
results in some unsteadiness as described above.
However, in the compressible case,
additional unsteadiness occurs because an associated 
unbalanced density change is generated. This was calculated and included
in the set up by noting that for a strictly steady state
horizontal flow  of the type considered here, 
$Q (\psi) =c_s^2\ln\rho +(1/2) {\bf u}^2 -(3/2)\Omega^2y^2 +\Phi_G,$
with $\Phi_G = \Omega^2z^2/2$ being the gravitational potential, 
should be a function only of the stream function $\psi.$
This is readily determined from the imposed steady flow
from the condition
\begin{equation}
\frac{dQ}{d\psi}= \frac{1}{2} \Omega + \omega_{v},
\end{equation}
where $\omega_{v}$ is taken to be the same function of $\psi$
as for the infinite vortex patch. The latter is of course an approximation
as we modified the infinite patch by imposing periodic boundary
conditions. Nonetheless its use to calculate 
the initial density
on horizontal plane resulted in initial states that became increasingly
more steady as $c_s$ increased.

In each of the simulations the two dimensional vortex structure
was found to remain for about $40$ shearing times before
undergoing a linear instability 
 which began with a very short wavelength in the vertical direction. The 
root mean square (weighted with density) vertical velocity is plotted
in Fig. \ref{jcbp1}. Initial values increase with $v_s/c$ because
   the initially imposed flow is further from being stationary.

Streamlines in the horizontal midplane are shown for
$v_s/c_s=1.3$ at $t= 37$,  and  $v_s/c_s = 0.13$ at $t=47$ in Fig. \ref{jcbp2}.
 Although linear instability has begun, 
 the initially imposed flow  is found to be  well preserved
 for the case with  $v_s/c_s = 0.13$ which most closely resembles the incompressible case.
 In this case the maximum growth rate in the linear phase of $0.75S$ is in good agreement
 with the linear prediction obtained  from equation (\ref{ling}).
 
   However, when $v_s/c_s = 1.3$ the flow
 shows significantly  greater time variability. The  vortex is squeezed
 in the radial direction and trailing  features begin to form outside
 the initial vortex core which turn into density waves which, however, are not well 
represented in our small computational domain.
 Global vortical structures survive until they are eventually destroyed
 by the instability
 Snapshots of the vertical velocity field in the horizontal midplane
  are  shown after
 the onset of linear instability in Fig.~\ref{jcbp3} for the three simulations.
 The 3D perturbations are found to be  localized inside the original elliptical
 boundary of the vorticity patch just as in the incompressible case.
     
\subsection{Stratified case}
\begin{figure}
\centering
\includegraphics[width=0.9 \linewidth]{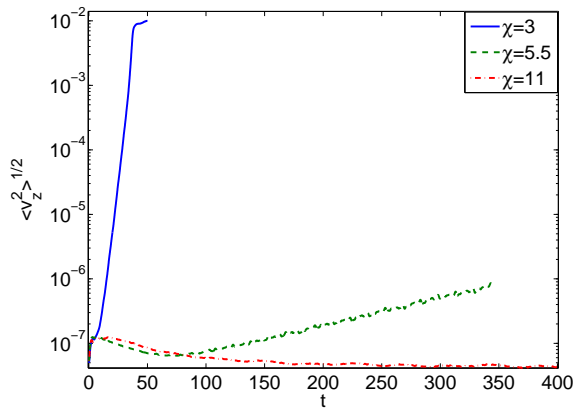}
 \caption{Time evolution of $\sqrt{\langle v_z^2\rangle}$ for several isolated vortices in stratified shearing box simulations with $N=S$. We find an exponential growth in agreement with the linear prediction for $\chi=3$ and $\chi=5.5$. The expected instability for $\chi=11$ is not observed, probably because of a too large numerical dissipation.}
              \label{vzenergN1}
\end{figure}

To study the vertically stratified case, we have implemented the Boussinesq equations (\ref{motiongeneralB}) in our spectral code. Compared to (\ref{motiongeneralB}), we have added an eighth order  hyper diffusivity 
to the velocity and potential temperature equations. 
We followed the same procedure as in the non stratified case to set up the vortex structure. 
We have run the simulations with a moderate stratification ($N=S$),
for which one expects a departure from the non stratified case  
when $\chi>4$ (Fig.~\ref{FloquetN1}). 
The time history of $\langle v_z^2\rangle^{1/2}$, for various vortex aspect-ratios, 
is presented in Fig.~\ref{vzenergN1}. 

For $\chi=3$ we find $\gamma=0.35$ and $\gamma=0.01$ for $\chi=5.5$. 
No instability is observed in our simulations for $\chi=11$.
These measured growth rates are always below the expected growth rate from our linear analysis
($\gamma_3=0.43$, $\gamma_{5.5}=0.024$ and $\gamma_{11}=0.004$) which might be due to excessive
numerical dissipation in the stratified case, 
despite the high resolution and the hyper viscosity prescription used. 
For both $\gamma=3$ and $\gamma=5.5$, the instability appears to be localised inside the vortex core, as expected.
 We can conclude from our results that the elliptical instability can be detected in stratified simulations,
 but resolution and numerical dissipation must be \emph{carefully} controlled to get accurate results. 

\section{Discussion}
We have described a 3D instability appearing in elliptical vortices (Kida vortices) embedded in accretion discs, 
known as the elliptical instability. We have shown analytically, {for the case of no stratification}
 that this elliptical instability is \emph{always} found in vortex cores, except for a narrow range of vortex aspect-ratio ($4<\chi<5.9$). 
When the vortex is weak (elongated vortex), 
the instability involves small radial wavelengths compared to azimuthal and vertical ones, the ratio being of the order of $\chi$. 
Moreover, the inclusion of a stable stratification tends to weaken the instability for large $\chi$ but it \emph{does not suppress it}.

Numerical simulations have essentially confirmed our linear analysis. 
We have found the predicted growth rate in the case where the instability was expected inside the core. 
Furthermore, instabilities \emph{outside} of the vortex core were observed when the core was linearly stable, 
in agreement with our physical interpretation.
 The inclusion of compressibility in the simulations didn't change the behaviour of the instability,
 and analytical results were recovered. We also studied numerically the effect of 
vertical Boussinesq stratification. Again, analytical results were recovered, except for weak vortices ($\chi=11$) for which stability was observed, probably because of numerical diffusion.

These results tend to indicate
that elliptical instabilities, or more generally parametric instabilities, 
are \emph{always} present in accretion disc vortices.
However, they often involve \emph{small radial wavelengths} (compared to the disc thickness) and small growth rates
(compared to the orbital time scale). Put together, these properties make elliptical
instabilities hard to capture numerically because of numerical diffusion at the grid scale. 
For example, highly accurate spectral methods with high-order hyper-diffusivity were required to marginally 
resolve the instability in a $\chi=11$ vortex. 

In this work, we have not addressed the saturation of this instability. 
This is however rather intentional, as we think that the non-linear outcome should be studied
 together with the production mechanism for  these vortices. In fact, the evolution of these structures will be dictated ultimately by the competition between the production mechanism and the saturated 3D instability. 
In this context, studying the saturation of this instability \emph{on its own} will be of little interest, 
since it will be modified by the vortex production process.
Note too that the time-scale needed by the saturated instability to destroy the vortex structure is unknown.
In particular, there is \emph{no reason} to think that this time-scale is related to the growth rate in the linear stage.
For example, a weakly unstable vortex could be destroyed in one orbit once saturation is reached.
Therefore, no firm conclusion can be drawn from this work concerning the evolution of the vortex structure \emph{itself}. 
However, in the absence of a sustaining production process, eventually 
one would expect the destruction of the vortex structure when the perturbation amplitude reaches 
the background flow amplitude as was indeed observed for our $\chi=3$ simulation.

This work has also several limitations. In particular, we have assumed an isolated 2D vortex as a starting point. 
However, one might want to study 3D vortices including a complicated vertical structure, with e.g.
 a mean vertical flow in the core (similar to cyclones on Earth for example).
 This kind of structure was suggested by \cite{BM05} for off-midplane vortices,
 and there is no doubt that stability properties will be modified in this case. 
Moreover, we have mostly restricted our study to elliptical streamlines, since these are tractable analytically.
 Although some extensions to more general cases were shown, we have not provided any formal proof
 of instability in the most general case. However, the  instability mechanism exhibited
 in the elliptical case appears to be quite general, and we think it's 
unlikely that a steady vortex could be stable \emph{everywhere} to 3D perturbations.

Finally, these results point out the necessity of carrying high-resolution 3D simulations of the mechanisms suggested in the literature for vortex formation. This includes for example Rossby wave instabilities \citep{LCWL01}, baroclinic instabilities \citep{KB03,PJS07} or gaps due to giant planets \citep{VAAP07}. In any case, the presence of 3D parametric instabilities will certainly modify the global outcome of these processes, with probably new physical effects.

\begin{acknowledgements}
      The authors thank Jeremy Goodman, Gordon Ogilvie, Orkan Umurhan and Fran\c cois Rincon for useful discussions. The simulations presented in this paper were performed using the Darwin Supercomputer of the University of Cambridge High Performance Computing Service (http://www.hpc.cam.ac.uk/), provided by Dell Inc. using Strategic Research Infrastructure Funding from the Higher Education Funding Council for England. GL acknowledges support by STFC . 
\end{acknowledgements}

\bibliographystyle{aa}
\bibliography{1577bib}

\end{document}